# Rydberg polaritons in ReS$_2$ crystals

A. Coriolano[1,2], L. Polimeno[1], M. Pugliese[1,2], A. Cannavale[1,3], D. Trypogeorgos[1], A. Di Renzo[1,2], V. Ardizzone[1,2], A. Rizzo[1], D. Ballarini[1], G. Gigli[1,2], V. Maiorano[1], A. S. Rosyadi[4], C.-A. Chuang[4], C.-H. Ho[4], L. De Marco[1], D. Sanvitto[1] and M. De Giorgi[1]

[1] CNR NANOTEC, Institute of Nanotechnology, Via Monteroni, Lecce 73100, Italy

[2] Dipartimento di Matematica e Fisica E. De Giorgi, Università del Salento, Campus Ecotekne, Via Monteroni, Lecce 73100, Italy

[3] Department of Civil Engineering Sciences and Architecture, Polytechnic University of Bari, Bari, Italy

[4] Graduate Institute of Applied Science and Technology, National Taiwan University of Science and Technology, Taipei 106, Taiwan

**Abstract**

Rhenium disulfide (ReS$_2$) belongs to group-VII transition metal dichalcogenide (TMDs) with attractive properties such as exceptionally high refractive index and significant oscillator strength, large in-plane birefringence, and good chemical stability. Unlike most other TMDs, the peculiar optical properties of ReS$_2$ persist from bulk to the monolayer, making this material potentially suitable for applications in optical devices. In this work, we demonstrate with unprecedented clarity the strong coupling between cavity modes and excited states, which results in a strong polariton interaction, showing the interest of such materials as a solid-state counterpart of Rydberg atomic systems. Moreover, we definitively clarify the nature of important spectral features, shedding light on some controversial aspects or incomplete interpretations and demonstrating that their origin is due to the interesting combination of the very high refractive index and the large oscillator strength expressed by these TMDs.

**Introduction**

Transition metal dichalcogenides with chemical formula $MX_2$ (e.g.: M= Mo, W, Re; X= S, Se) are Van der Waals (vdW) materials with outstanding structural and optical properties, such as chemical stability[1], mechanical flexibility[1], high binding energies[2,3] and oscillator strengths, and narrow photoluminescence linewidths[4], which make them extremely attractive in a plethora of photonics and optoelectronics applications[5–7]. Their optical and chemical properties vary according to which group the TMDs belongs.

In particular, group VI TMDs, such as $MoS_2$, $MoSe_2$, $WS_2$, and $WSe_2$ are characterized by linear isotropic in-plane optical properties due to the high symmetry of their crystal structure. Moreover, they show a transition from indirect to direct bandgap when going from bulk to a monolayer. This is due to their strong interlayer coupling, which is broken when the out-of-plane confinement is achieved with single layers[8].

On the other hand, group VII TMDs, such as $ReS_2$ and $ReSe_2$, crystallize in a distorted single-layer trigonal (1T') structure of triclinic symmetry (Fig. 1a) due to the Re-Re interaction aligned along the b-axis. This results in reduced crystal symmetry, that leads to strong in-plane anisotropic optical properties[9,10], inducing the formation of two almost orthogonally polarized in–plane excitons[9] and high optical birefringence[11,12]. These properties are exploited for different applications, such as field effect transistors[13,14], polarized photodetectors[13,15], and photocatalyst[16]. Unlike other TMDs, $ReS_2$ and $ReSe_2$ are also characterized by a direct bandgap that persists from bulk to monolayer, due to the distorted 1T structure[17] that hinders ordered stacking of neighboring layers and minimizes the interlayer overlap of wavefunctions, as shown by density functional theory calculations[18]. Such a weak interlayer coupling makes it possible to achieve the same properties as two-dimensional systems, regardless the number of layers, avoiding the challenging and time-consuming preparation of large-area monolayers. In addition, they have a very high refractive index in the visible/near infrared spectral region, a quite unique feature compared to other materials[19]. This makes $ReS_2$ extremely interesting for photonic applications and a unique platform for the exploration of novel topological properties when used as metamaterials.

Materials with planar optical anisotropy that support matter-light quasiparticles (i.e. polaritons), resulting from the strong coupling between excitons and photons, are extremely interesting thanks to their potential for the realization of topological exciton-polariton systems. This is mainly due to the possibility of easily tuning the optoelectronic properties of the polariton device, by actively changing different parameters, such as crystal thickness, polarization, external magnetic and electric field[20] and sample temperature. Moreover, it has been theoretically predicted that exciton-polariton condensates can exhibit longer-range algebraic correlations under non-equilibrium conditions only in strongly anisotropic systems[21]. All these reasons make the highly anisotropic $ReS_2$ crystals, as active materials in exciton-polariton systems, very intriguing[22,23].

In this work, we unambiguously demonstrate the polarization-dependent strong coupling in $ReS_2$ crystal and the hybridization between different higher order exciton states resulting in Rydberg polaritons with enhanced interactions, making this material useful for the realization of polarisation-controlled polaritonic devices. Taking advantage of the spectral features of strongly coupled $ReS_2$ polaritons we clearly demonstrate that $ReS_2$ crystals possess only two orthogonally-polarized excitons confuting previous studies suggesting four excitonic resonances[24,25].

**Experimental details**

$ReS_2$ bulk crystals were grown by chemical vapor transport (CVT) method using $ICl_3$ as the transport agent[9]. Due to the weak Van der Waals bonding between the layers, micrometer-sized $ReS_2$ flakes with desired thickness were obtained by mechanically exfoliating from bulk crystals and transferring them onto the final substrate by a dry transfer method[26] using commercial polydimethylsiloxane (PDMS). The flakes were transferred on glass or Distributed Bragg Reflectors (DBR) formed by 8-pairs of $SiO_2/TiO_2$ with a stopband centered at $\lambda_c = 785$ nm, grown on glass substrate by electron-beam deposition (see Supporting Information for details). By using a closed-cycle cryostat (Attodry1000), the sample is cooled down to liquid helium temperature (T = 4 K) and excited by a white halogen lamp, in order to measure the reflection spectra both in real and Fourier space. The polarized spectra along the b-axis (V-polarized, $\phi = 0$) and perpendicular to it (H-polarized, $\phi = 90$) were obtained using a half wave plate and a polarizer placed in front of a spectrometer coupled to a charge-coupled device (CCD).

**Results and discussion**

The $ReS_2$ crystal structure is shown in Fig.1a: each crystal layer of Re atoms is placed between two S sheets, with distorted trigonal anti-prismatic coordination and strong covalent bonding between the Re and S atoms. Rhenium atoms (grey) form a chain due to the Re-Re bonds which defines the b-axis of the crystal. Owing to the strong metal-metal bond, $ReS_2$ breaks preferentially along the b-axis[27,28], typically forming a longer crystal edge after mechanical exfoliation.

The distortion of the ReS$_2$ atomic structure induces the strong anisotropy of the exciton resonances, resulting in a different orientations and optical selection rules for linearly polarized light[25,29,30]. In fact, the typical polarized reflection spectrum for a 35 nm thick ReS$_2$ crystal exfoliated on glass substrate (top layer of Fig. 1b) are mainly characterized by one exciton resonance, $E_{x1}$, polarized parallel to the b – axis ($\phi = 0°$) (black line), and a second exciton resonance, $E_{x2}$, polarized almost perpendicular to the b – axis ($\phi = 90°$) (red line). Moving to thicker crystals we observe some changes in the reflection spectra: 80 nm-thick ReS$_2$ exhibits an exciton resonance splitting for each polarization and excited states transitions appear at higher energy (~1.65 eV) as shown in the bottom layer of Fig.1b.

Additional resonances around the main exciton have already been observed in previous works[22,23], and as in our case, these features have been clearly distinguishable for crystals exfoliated in thick flakes ($\geq$ 50 nm)[9,25], whereas for the thinner one only the main exciton transition is distinctly observable[9,30–32]. However, their nature is still debated. In the work of Arora et al.[25], multiple close-lying bright excitons were associated to twofold degenerate direct transitions from the valence band maximum to the conduction band minimum, with each degenerate pair consisting of bands with opposite spins. Additional peaks appearing below the exciton in thick crystals have been attributed to donor bound excitons[9], whereas Dhara et al.[23,24] hypothesized that these peaks are i) due to the splitting of singlet and triplet states of excitons as a result of the electron-hole exchange interaction or ii) induced from the broken rotational symmetry due to the structural anisotropy and spin-orbit coupling of ReS$_2$. Recently, R. Gogna et al.[22] hint at an apparent splitting of the exciton resonance due to the cavity effect caused by reflections within the flake. Herein, we clearly assess that the two resonances that appear for both polarized excitonic transitions are in fact the longitudinal and transverse excitons, to which the polaritonic branches asymptotically tend and observable in ReS$_2$ crystals for flakes thicker than 50 nm. This is due to the unique combination of material optical parameters, mainly i) the very high background refractive index and ii) the large oscillator strength associated to the ReS$_2$ excitons.

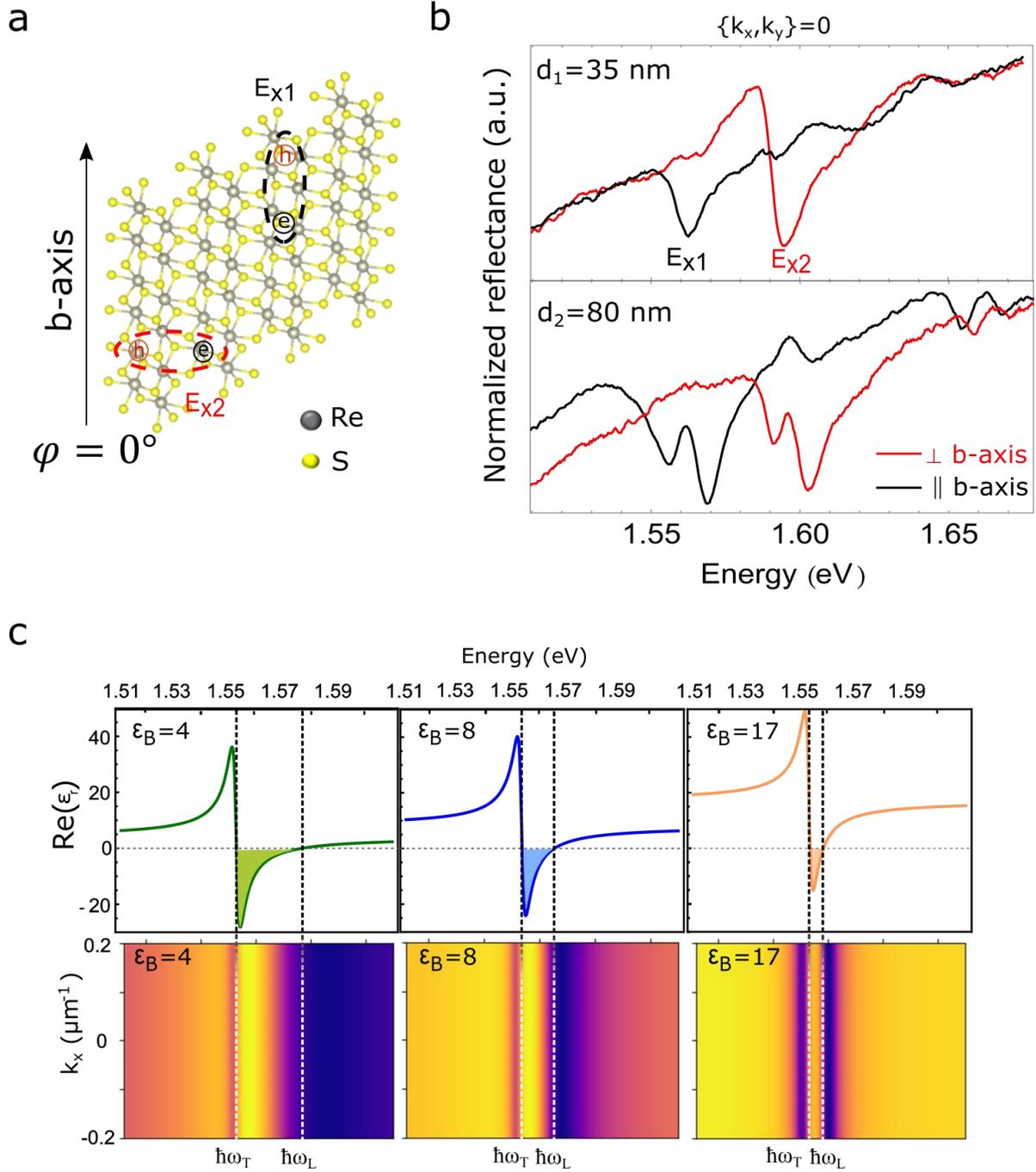

Figure 1. (a) Sketch of ReS$_2$ monolayer atomic arrangement. The Re4 clustering chains, b-axis, and the dipole direction of the main excitons are also indicated. The angle ϕ is defined as the polarization angle with respect to the b-axis. (b) Reflection spectra of a 35 nm (top panel) and an 80 nm (bottom panel) thick ReS$_2$ crystal exfoliated on a glass substrate for parallel (black line) and perpendicular (red line) polarization respect to the b-axis. (c) Top panels show how different values of the real part background permittivity, ε$_B$, affects the reflectivity spectra close to the exciton resonance. Note that the most left and right panels reproduce the effect of a low and high permittivity on materials with a high oscillator strength, such as perovskites and ReS$_2$ respectively.

To investigate the role of the high background refractive index on the reflectivity of the ReS$_2$, we consider the reflectance, at zero-order approximation, given by $R = \frac{(1-n)^2 + k^2}{(1+n)^2 + k^2}$ where n and κ are the real and imaginary part of the complex refractive index, respectively. Due to the very high real part of the refractive index, $n = \sqrt{Re[\varepsilon(\omega)]} > 4$ (where ε(ω) is the complex permittivity), the crystal flakes, when lying on a low index material, behave as a dielectric slab resonator supporting Fabry Perot modes. Thanks to the high oscillator strength, f ~ 0.3 eV$^2$, of the excitonic transitions, there is a strong interaction between these modes and the exciton resonances giving rise to new hybrid states called polaritons[33].

This can be easily described by modelling the two polarization-dependent exciton resonances, $E_{X1}$ and $E_{X2}$, with a Lorentz oscillator, using a dielectric function given by:

$$\varepsilon_{1,2}(\omega) = \varepsilon_{B_{1,2}} + \frac{f_{1,2}}{E_{x_{1,2}}^2 - E^2 - iE\Gamma_{1,2}} \qquad (1)$$

where $E = \hbar\omega$, $\varepsilon_B$ is the background permittivity, $E_{X1,2}$ are the two excitonic resonances, Γ is the exciton linewidth, and $f$ is the oscillator strength. We found that in ReS$_2$ due to the peculiar combination of high refractive index and strong exciton oscillator strength, the real part of the permittivity crosses zero and becomes negative around the exciton resonance (Fig. 1c top right panel). By considering the spatial evolution of the field in this negative epsilon region (negative permittivity), following Maxwell's equations[34] we can obtain two class of solutions given by i) ε(ω) = 0 and ii) $\vec{k} \cdot \vec{E}_{ElectricField} = 0$. The first solution $E_L = \hbar\omega_L$ corresponds to the appearing of a longitudinal mode, which is usually invisible in ordinary materials, while the second solution $E_T = \hbar\omega_T$ corresponds to the standard transversal mode[35,36]. As a result of the appearance of both self-hybridized modes, in the region with $Re[\varepsilon] < 0$ the electromagnetic wave cannot propagate into the material, but rather exponentially decay, resulting in a strong effective reflectivity[37]. It is worth nothing that the energy position of the longitudinal mode and so the gap between $E_L$ and $E_T$ strongly depends on the background permittivity of the material, $\varepsilon_B$. By decreasing $\varepsilon_B$ the longitudinal modes shift at higher energy, resulting in an increase of the longitudinal-transversal energy splitting $\Delta E_{L/T} = \hbar\omega_L - \hbar\omega_T$. (See top panel Fig.1c) but with a smoother transition between the region with $Re[\varepsilon] < 0$ and $Re[\varepsilon] > 0$. On the other hand, for fixed background permittivity, $\varepsilon_B$, $\Delta E_{L/T}$ increases for higher coupling strength between the exciton and photons (i.e. higher oscillator strength of the exciton resonance, Fig. S1 of Supporting Information). It is therefore clear that both modes can be sharply seen only in those materials that possess a high permittivity while keeping an equally high oscillator strength. ReS$_2$ has the chance to meet both criteria.

The polarized reflectance spectrum of ReS$_2$ crystals exfoliated on top of a glass substrate has therefore such characteristic. Simulations using the semi-analytical Rigorous Coupled-Wave

Analysis (RCWA) method[38] depict very well this behavior. It is interesting to note, looking at Fig. 1c, that materials such as perovskites, (parameters simulated in the bottom left panel) with a high background refractive index (n ~ 2, i.e. $\varepsilon_B$ ~ 4) and a real part of the permittivity that also becomes negative, do not show two sharp resonances due to the smooth variation of the permittivity if compared to the $ReS_2$, which is simulated in the righthand side of Fig. 1c.

In the following, we exploit the high refractive index and the strong oscillator strength of the excitonic transitions in $ReS_2$ to investigate the full hybridized dispersion of the ground state excitons and the first two excited states under strong coupling regime. In order to do so we have exfoliated a 310 nm-thick $ReS_2$ crystal (Fig. 2a and S3) on a distributed Bragg reflector (DBR) that would increase the mode finesse without reducing excessively the PL collection from the front side of the crystal. Fig. 2b-c shows the linearly polarized energy reflection as function of the in-plane momentum, $k_x$. The spectra unambiguously evidence the typical dispersions of a system in strong light-matter coupling regime, with the folding of the energy bands for the two excitons depending on the direction of polarisation.

Because of the strong in-plane optical birefringence present in $ReS_2$ crystal, the real part of the refractive index along the Re-Re chain direction (b-axis, $\phi = 0°$ n = 4.1) is ~20% higher than the ones along the perpendicular direction ($\phi = 90°$, n = 3.2) [11,39,40]. Consequently, the photonic modes of the structure shift at higher energy for the H polarization with respect to the V polarization. This results in a different detuning of the polariton states with a different fraction of photon and exciton for each polaritonic band in the two linear polarizations.

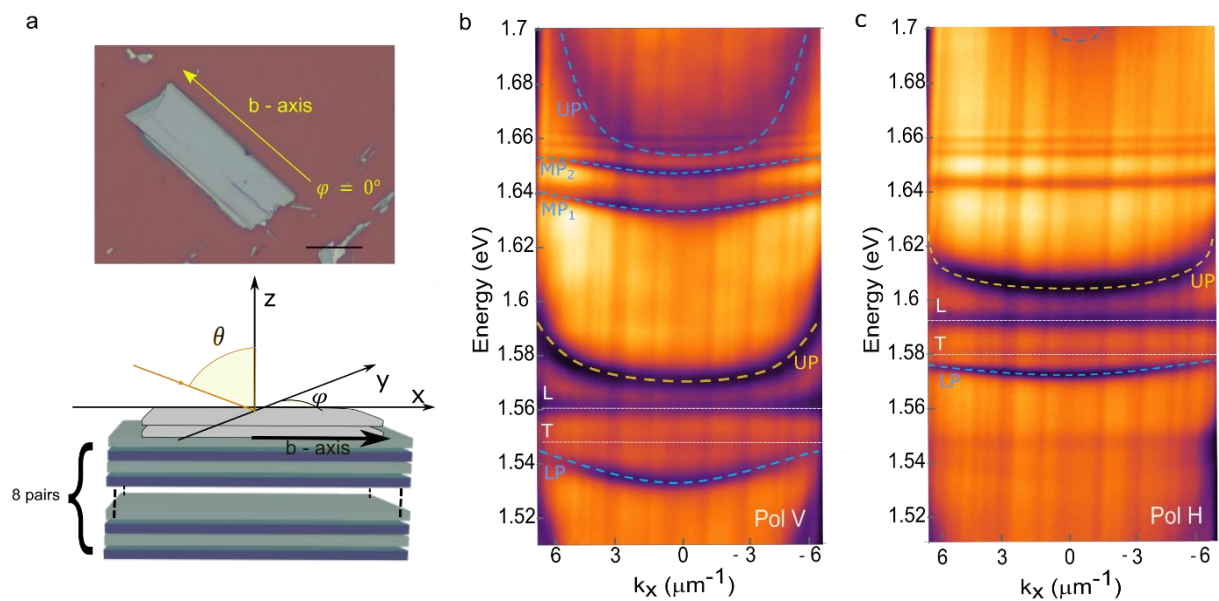

Figure 2. (a) Top panel shows the optical microscope image of a 310 nm thick ReS$_2$ flake exfoliated on a DBR. The scale bar is 20 $\mu m$. The bottom panel is a scheme of the structure composed of a bottom DBR made by 8 SiO$_2$ / TiO$_2$ pairs with a ReS$_2$ crystal (gray) on top. (b-c) Energy vs k$_x$ in-plane momentum reflection spectra polarized parallel to the b – axis (a) and perpendicular to the b – axis (b). The blue dashed lines are the calculated polariton branches resulting from the strong coupling of the exciton with a cavity mode; the yellow line is the polariton upper polariton (UP) branch induced by the coupling of the exciton with the previous mode, whereas the white lines are the energies of the transversal (T) and longitudinal (L) modes.

A detailed theoretical analysis of the reflectivity at {k$_x$, k$_y$} = 0 as a function of the crystal thickness has allowed to associate the various dispersion curves for the two polarizations. For the V polarization, both the exciton E$_{X1}$ and the excited states are strongly coupled to a Fabry-Perot photonic mode (Fig S4a), resulting in the formation of lower (LP, at 1.533 eV), middles (MP$_1$, at 1.632 eV and MP$_2$, at 1.647 eV) and upper polariton (UP, at 1.653 eV) branches (dashed blue guidelines and label in Fig. 3c). The polariton state evidenced by the dashed yellow line instead represents the upper polariton branch, generated by the strong coupling of the exciton E$_{X1}$ with a previous Fabry Perot photonic mode (energy of the bare mode ~1.4 eV). The corresponding lower polariton branch of this other Fabry Perot mode is not experimentally observable because its energy is outside the stop band of the DBR. Finally, the transitions evidenced by the pointed dashed white lines are the transversal (T) and longitudinal (L) modes as described in figure 1c.

The dispersion spectrum in the H polarized direction (Fig. 3d) shows a similar behavior for the exciton E$_{x2}$ despite a different detuning of the mode and lower oscillator strength. However, the excited states remain uncoupled being the photonic modes highly detuned from these energies (see also Fig S3b). By changing the ReS$_2$ crystal thickness, we can also be able to tune the coupling of the exciton and the one of the excited states varying the number and position of the photonic modes (Fig. S5 and S6).

By making a 1D raster scan we can reconstruct the reflectance spectra in both direction of the Fourier plane. The k$_x$-k$_y$ dispersion maps for the two polarizations are plotted in figure 3a in terms of the degree of polarization (DOP), defined as (I$_V$-I$_H$)/(I$_V$+I$_H$), where I$_H$ and I$_V$ are the reflected light intensities for horizontal and vertical polarizations, respectively.

Figures 3a and 3e demonstrate the good agreement between the experimental DOP (Fig. 3a) along the k$_x$ direction at k$_y$ = 0 with the theoretical calculations (Fig. 3e). While modes that are well separated in energy show isotropic polarization in k$_x$ and k$_y$ (see in Fig. 3b, c, f and g), at the energy where modes of different polarization cross each other at high k vector, k$_x$ ~ ± 6 µm$^{-1}$, the DOP shows a polarization rotation along the dispersion in k-space (see Fig.3 d and h)

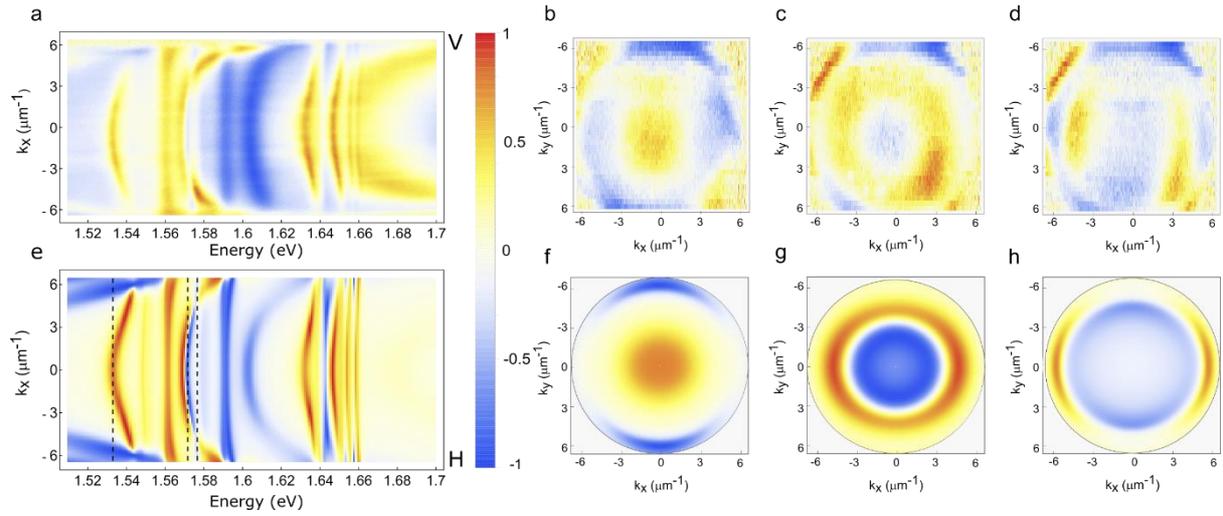

Figure 3. Experimental (a) and theoretical (e) DOP of the reflectance signal, in the Horizontal/Vertical (H/V) basis. Experimental (b-d) and theoretical (f-h) degree of linear polarization of the reflectance maps in the momentum space at isoenergetic cross-sections corresponding to the lower polariton of the exciton $E_{X1}$(b, f), the lower polariton of the exciton $E_{X2}$ (c, g) and at the energy where the lower polariton of exciton $E_{X2}$ cross at the higher $k_x$ values the upper polariton of $E_{X1}$ (d, h).

The excited states, clearly visible in Fig. 2b above 1.64 eV, also show strong coupling with same slab-cavity mode. The values of the Rabi and coupling strength for each exciton state are extracted by fitting our dispersions with a four coupled oscillators model (one mode, the ground exciton state and two excited states). The result of this fitting is shown in Fig S7a with an extracted Rabi splitting of 84 meV, 20 meV and 12 meV for the lower and the middles polariton states, respectively. By focusing on the Hopfield coefficients which give the contribution of the excitons and the photon mode to the different polariton states, we can estimate the character of the polariton modes and how the polariton-polariton interaction should change in each of the three polariton branches. In Fig. 4a, which shows the Hopfield coefficients calculated for the lower and the first middle polaritonic bands, it can be seen that the lowest exciton-polariton mode (left panel of Fig. 4a) is practically decoupled from the excited excitons, and the state is predominantly having a ground state character (with 80% of the fundamental exciton and only a 20% of photon contribution). Surprisingly, the first middle polariton branch (right panel of Fig. 4a) is composed of three excitonic components, the fundamental and the two excited excitons. We can speculate then that such a state should lead to stronger polariton nonlinearity described with an interaction coefficient, $g_{P-P}$ enhanced by the presence of higher order excitons[41,42]. In fact, for the excited states (called also Rydberg states) the radius of the different orbits scales as $n^2$. Since the exciton–exciton interaction $g_{X-X}(n)$ depends linearly on the Bohr radius ($a_B$), its intensity should increase quadratically: $g_{X-X}(n) = 6E_b(n)a_B^2 n^2$, where

$E_b(n)$ is the binding energy of the n$^{th}$-state extracted by fitting the energy position of the different states in the absorption spectrum (Fig. S8). Since the polariton-polariton interaction is given by the weighted sum of the different exciton contributions $g_{p-p} = \sum_n \gamma_{X_n}^2 g_{X-X}(n)$ where $\gamma_{X_n}$ represents the normalized Hopfield coefficient for the n$^{th}$-state, it should be proportionally higher for the middle branch which have higher content of the Rydberg excitons. In order to experimentally observe such polariton-polariton interaction, we resonantly excite the ReS$_2$ crystal with a very fast, broadband, pulsed laser (~ 35 fs) polarized along the b-axis.

We observe that whereas the reflection spectrum of the lower polariton state does not change by increasing the excitation power (left panel of Fig. 4b), possibly due to the limited nonlinearity of the ground state exciton, the spectrum of the first middle polariton branch show a small shift towards higher energies (right panel of Fig. 4b) [43,44]. The energy blueshift of the reflected peak is plotted in Fig. 4c for the middle polariton state against the excitation power resulting in a small but observable blueshift of about 1 meV. Although the second middle polariton should be characterized by a stronger polariton-polariton interaction we do not observe any spectral shift with the pumping power. In general, we believe that due to the very low quantum efficiency of this material[45] the real power density injected under resonant excitation is extremely low. This could be also the reason for such a small blueshift of the middle polariton branch and the lack of observable shift for the ground state polariton.

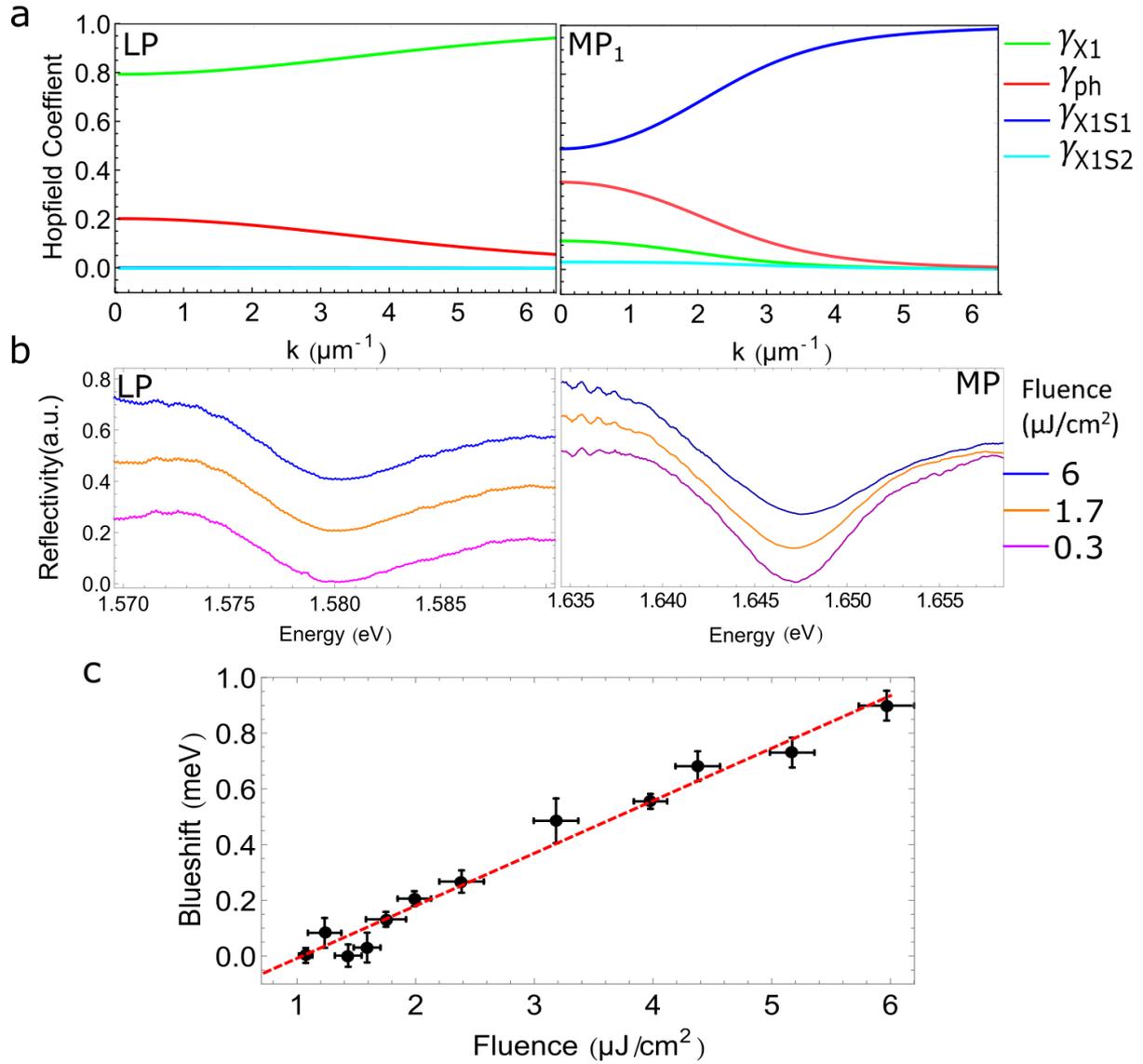

Figure 4. (a) Hopfield coefficients extracted for the lower (left panel) and first middle (right panel) polariton branch. (b) Reflection spectra obtained by resonantly exciting the lower and the middle polariton branches at k=0 for different excitation pump powers. (c) Blueshift of the middle polariton by resonantly exciting the sample with a laser linearly polarized to the b-axis; the dashed red line is the linear fit to the experimental data with a slope of 0.18 meV cm$^2$/μJ.

**Conclusions**

In conclusion, we investigated the effect of the high refractive index and strong oscillator strength of the exciton transitions in ReS$_2$ crystals and clearly elucidated it.

By hybridizing the ground state excitons and the two first excited states we were able to measure the effect of exciton-exciton interaction on the Rydberg states, showing a much stronger polariton-polariton interaction compared to the lower energy state. Moreover, we have definitively demonstrated that this material possesses only two orthogonally-polarized excitons and we attributed the additional line in the reflectivity spectra to the longitudinal exciton feature. This work, beyond shedding light on the presence of different transition lines previously observed in ReS$_2$ but mistakenly understood, provides new insights into this exciting material that could be successfully implemented in the realization of optical devices that exploit higher order Rydberg states in solid-state materials.

**Acknowledgement**

The authors gratefully thank S. Carallo, T. Stomeo and G. De Marzo for AFM measurements, P. Cazzato for technical support, F. Todisco and L. Dominici for useful discussions. This work was supported by the Italian Ministry of University (MIUR) for funding through the PRIN project "Interacting Photons in Polariton Circuits" — INPhoPOL (grant 2017P9FJBS), the project "Hardware implementation of a polariton neural network for neuromorphic computing" – Joint Bilateral Agreement CNR-RFBR (Russian Foundation for Basic Research) – Triennal Program 2021–2023, the MIUR project "ECOTEC - ECO-sustainable and intelligent fibers and fabrics for TEChnic clothing", PON « R&I» 2014–2020, project N° ARS01_00951, CUP B66C18000300005, the MAECI project "Novel photonic platform for neuromorphic computing", Joint Bilateral Project Italia-Polonia, 2022-2023 and the project "TECNOMED - Tecnopolo di Nanotecnologia e Fotonica per la Medicina di Precisione", (Ministry of University and Scientific Research (MIUR) Decreto Direttoriale n. 3449 del 4/12/2017, CUP B83B17000010001). C.-H. H. thanks the funding support from Ministry of Science and Technology, Taiwan under MOST 110-2112-M-011-002.

# Supporting Information: Rydberg polaritons in ReS2 crystals

A. Coriolano[1,2], L. Polimeno[1], M. Pugliese[1,2], A. Cannavale[1,3], D. Trypogeorgos[1], A. Di Renzo[1,2], V. Ardizzone[1,2], A. Rizzo[1], D. Ballarini[1], G. Gigli[1,2], V. Maiorano[1], A. S. Rosyadi[4], C.-A. Chuang[4], C.-H. Ho[4], L. De Marco[1], D. Sanvitto[1] and M. De Giorgi[1]

[1] CNR NANOTEC, Institute of Nanotechnology, Via Monteroni, Lecce 73100, Italy

[2] Dipartimento di Matematica e Fisica E. De Giorgi, Università del Salento, Campus Ecotekne, Via Monteroni, Lecce 73100, Italy

[3] Department of Civil Engineering Sciences and Architecture, Polytechnic University of Bari, Bari, Italy

[4] Graduate Institute of Applied Science and Technology, National Taiwan University of Science and Technology, Taipei 106, Taiwan

**ReS$_2$ crystal growth and sample preparation**

Layered single crystals of ReS$_2$ were are grown by chemical vapor transport (CVT) using ICl$_3$ as the transport agent, as reported in a previous work[1]. Thanks to weak Van der Waals force between the layers, few-layers of ReS$_2$ are obtained by mechanical exfoliation using commercial PDMS (Gel Pak) and transferred on glass or DBR substrates. Due to the in-plane mechanical anisotropy along the Re chains, the crystal b-axis is identified as a longer crystal edge, as shown in Fig. 2a of the main text, obtained after the mechanical exfoliation. To improve the crystal adhesion on the substrate and optimize the flakes quality, we pre-treated the surface of the DBR substrate with 4-(2-aminethyl) benzoic acid in order to create a self-assembled monolayer (SAM): the substrate was immersed in a 0.13 %wt solution of 4-(2-aminethyl) benzoic acid hydrochloride in ethanol for 24 hours at room temperature and then rinsed with ethanol to remove excess molecules.

**Distributed Bragg Reflector**

The DBR is formed by eight pairs of SiO$_2$/TiO$_2$ layers (with thicknesses of 136 nm/93 nm, respectively), deposited by electron-beam deposition (Temescal Supersource) in vacuum, keeping the chamber at $10^{-5} \div 10^{-6}$ mbar throughout the process, at room temperature (deposition rates: 1 Å/s for SiO$_2$, 0.5 Å/s for TiO$_2$). The DBR is deposited on top of a 170 μm glass substrate and the resulting stopband is centered at 785 nm.

**Optical measurements**

All optical measurements are performed in reflection configuration in a helium cryostat at cryogenic temperature (T=4K) using a halogen white lamp. The photoluminescence is recorded in reflection configuration with a continuous-wave laser, centered at λ = 488 nm. The spectra are collected with an 50x objective, with a numerical aperture NA=0.82. The polarizer and the half-wave plate (HWP) on the detection path allow for the measurement of the polarization-resolved spectra, both in real space and in back focal space.

**Theoretical Simulations**

Our structure has been simulated by using the semi-analytical Rigorous Coupled-Wave Analysis (RCWA) method implemented by S4 package[2]. The exciton resonances, $E_{X1}$ and $E_{X2}$, have been modeled using a Lorentz oscillator, with a dielectric function given by:

$$\varepsilon_{1,2} = \varepsilon_{B_{1,2}} + \frac{f_{1,2}}{E_{x_{1,2}}^2 - E^2 - i E \Gamma_{1,2}}$$

where:

$\epsilon_B$ is the background permittivity

$E_X$ is the exciton resonance energy

$\Gamma$ is the exciton linewidth

$f$ is the oscillator strength

After the measurement of the ReS$_2$ crystals thickness by AFM, we simulate the experimental reflectance spectra by using the following parameters, summarized in Tab. S1.

|  | $\epsilon_B$ | $f$ [eV²] | $\Gamma$ [eV] |
|---|---|---|---|
| ***Polarization // to b-axis (V)*** | 17 | $f_{x1} = 0.3$ | $\Gamma_{x1} = 0.003$ |
|  |  | $f_{S1} = 0.03$ | $\Gamma_{S1} = 0.0025$ |
|  |  | $f_{S2} = 0.022$ | $\Gamma_{S1} = 0.0025$ |
| ***Polarization ⊥ to b-axis (H)*** | 13.5 | $f_{x2} = 0.25$ | $\Gamma_{x2} = 0.003$ |
|  |  | $f_{S1} = 0.018$ | $\Gamma_{S1} = 0.002$ |

*Table S1. Values of the parameters utilized in the RCWA simulations.*

**Theoretical Simulations of ReS$_2$ crystal in vacuum**

Fig. S1 shows the real part of the permittivity by fixing the value of background permittivity, $\varepsilon_B$, for different oscillator strength ($f$). We consider a 60 nm – thick $ReS_2$ crystal immersed in vacuum and simulate the energy dispersion of the reflectivity for different values of oscillator strength. The energy splitting $\Delta E_{L/T}$ increases for higher coupling strength between the exciton and photons.

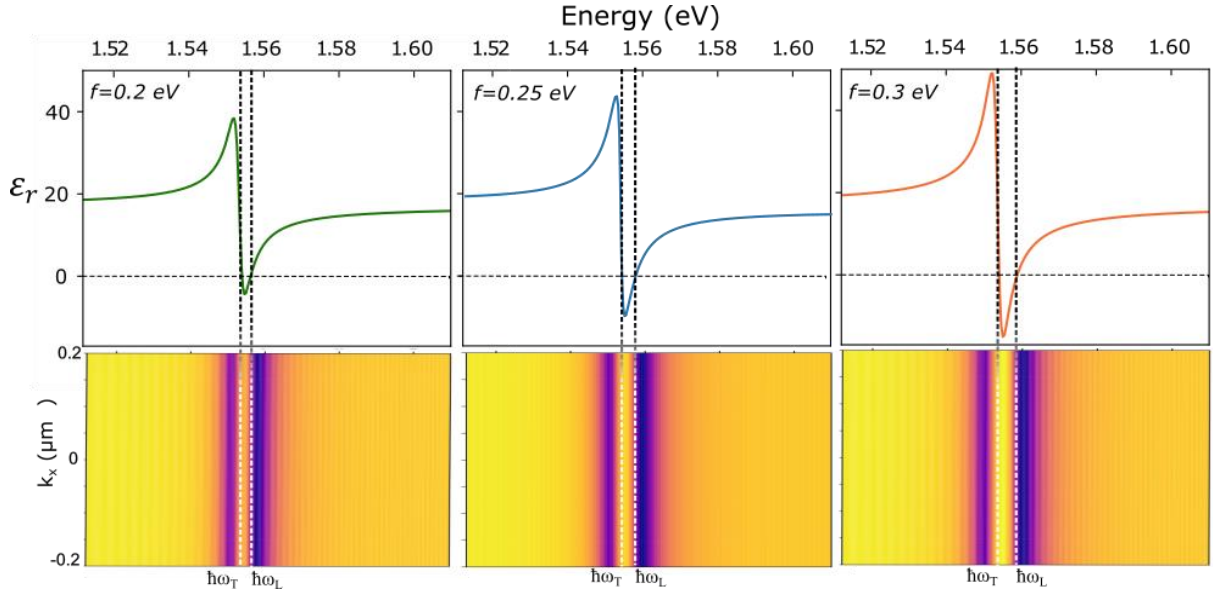

*Figure S1. (Top panels) Real part of the permittivity for different values of oscillator strength. The correspondent energy dispersions of the reflectivity for different values of f are reported in the bottom panels.*

**Theoretical simulation of ReS₂ reflectivity in function of the thickness**

The reflectivity spectra for different thickness of $ReS_2$ crystals on glass substrate are reported in Fig. S2. Additional resonances around the main exciton are clearly distinguishable for thick crystals ($\geq$ 50 nm), whereas for the thinner one only the main exciton transitions are distinctly observable.

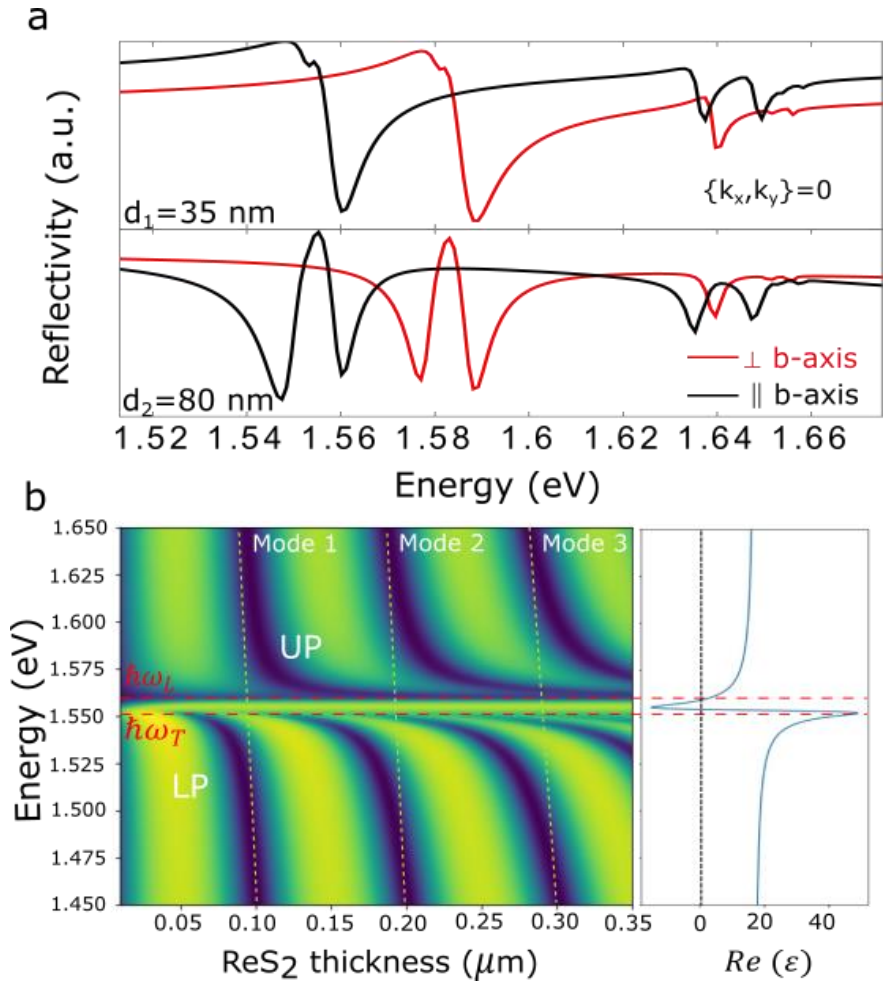

*Figure S2. (a) Theoretical reflectivity spectra of crystals with thickness of 35 nm (Top panel) and 80 nm (Bottom panel). (b) (Left panel) Theoretical map of the reflectivity versus crystal thickness for vertical polarization assuming only an exciton resonance at 1.554 eV and unpolarized light incident perpendicular to the surface crystal. The dashed red lines are the longitudinal and transversal modes for the $ReS_2$ crystals which couple to the exciton resulting in a new hybrid states, lower (LP) and upper (UP) polariton bands tending asymptotically at the longitudinal (L) and transversal (T) modes. (Right panel) Real part of the theoretical permittivity in the energy range of the reflectivity map.*

**Atomic Force Microscopy (AFM)**

AFM measurements are performed using a Park Scanning Probe Microscope (PSIA), with a high resonant frequency non-contact cantilever. The image acquisition is performed in air at room temperature.

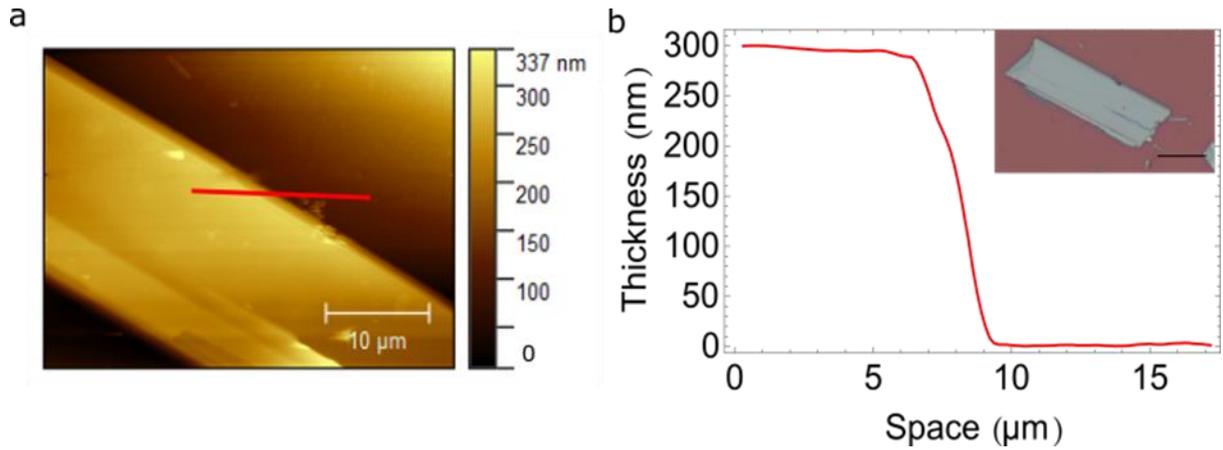

*Figure S3. (a) AFM topography image of 310 nm – thick ReS$_2$ crystal on DBR. (b) The corresponding height profile measured along the red line. (Inset) Optical image of the crystal. The scalebar is 20 µm.*

## Theoretical Simulations of the ReS$_2$ structure on top of DBR

We simulate the reflectance spectra at $\{k_x, k_y\} = 0$ for different thickness of the ReS$_2$ crystals on DBR, showing a different coupling effect for the two polarizations, in particular on the excited states.

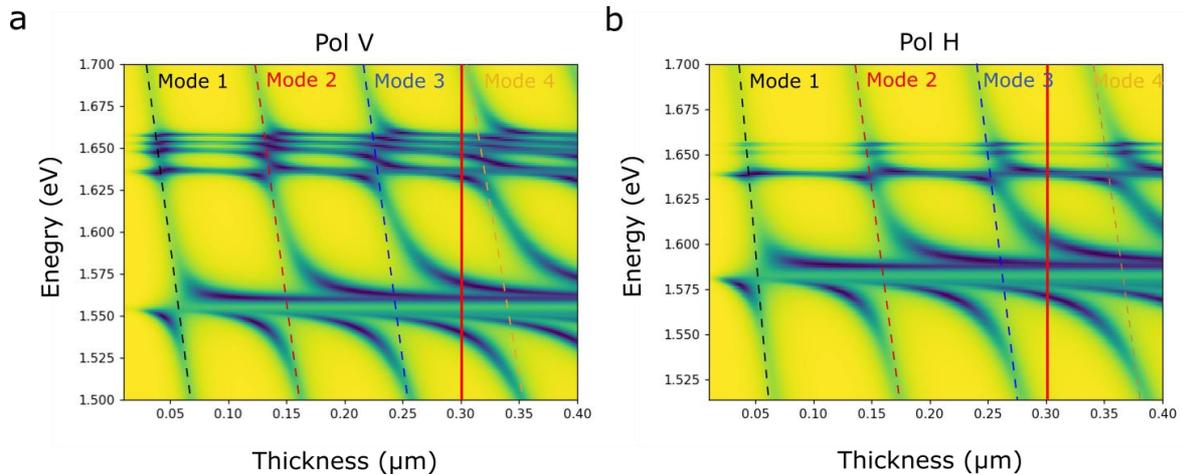

*Figure S4. Theoretical map of the reflectivity versus crystal thickness for the vertical (a) and horizontal polarization (b) assuming unpolarized light incident perpendicular to the surface crystal. The vertical red lines mark the energy dispersions for both polarizations for a crystal 310 nm thick, in agreement with the experimental data reported in Fig. 2a-b.*

## Tuning of ReS$_2$ crystal thickness

By changing the ReS$_2$ crystal thickness, we tune the coupling of the exciton and the excited states with different photonic modes: in a 60 nm-thick crystal, both excitons couple with the photonic mode 1, leading to the formation of the lower polariton states and the middle ones, while the excited states are uncoupled for both the polarizations (see Fig. S5).

For thicker sample (d = 140 nm), we can measure the coupling of the excitons with both the photonic Mode "2" and Mode "1", resulting in a more structured dispersion spectrum (see Fig. S6). Analyzing the V polarization spectra for different thickness, we conclude that the different detuning of the photonic modes with respect to the exciton $E_{X1}$ results in a different exciton component of the lower polariton branches, which becomes more excitonic from d= 310 nm to d = 60 nm (see Fig. S7).

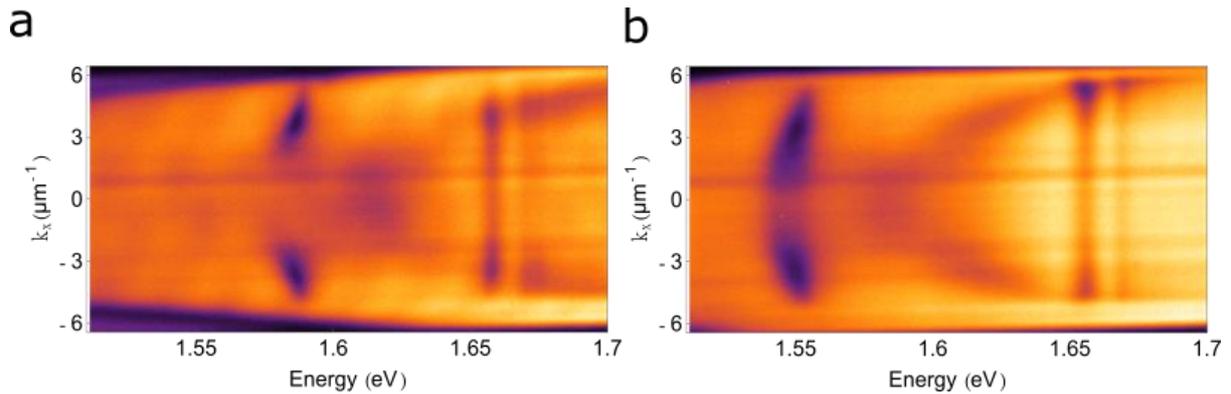

*Figure S5. Energy vs kx in-plane momentum of reflection spectra polarized perpendicular to the b – axis (a) and parallel to the b – axis (b) for a 60 nm – thick ReS$_2$ crystal on a DBR .*

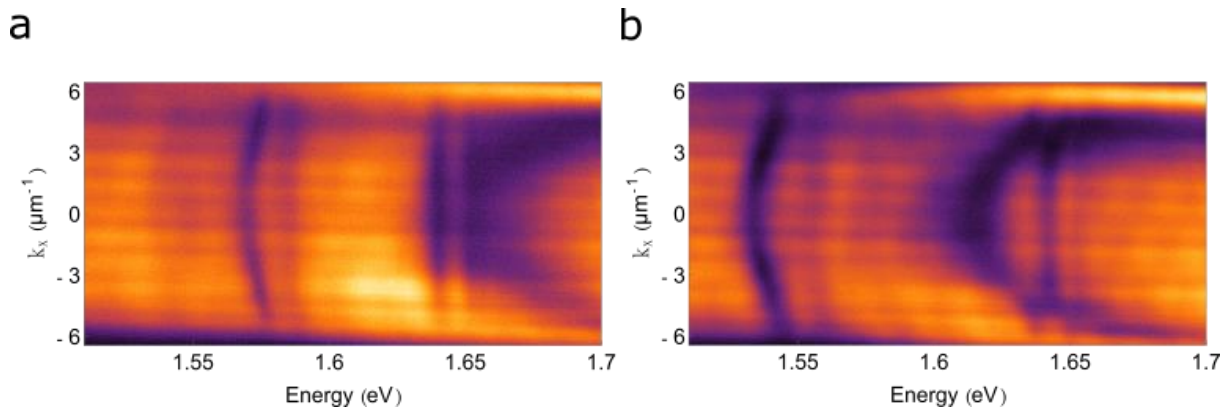

*Figure S6. Energy vs kx in-plane momentum of reflection spectra polarized perpendicular to the b – axis (a) and parallel to the b – axis (b) for a 140 nm – thick ReS$_2$ crystal on a DBR.*

To extract the Rabi splitting values from reflection spectra reported in Fig. 2b-c of the main text, we diagonalize a four coupled oscillator system matrix, described as follow:

$$\begin{pmatrix} E_{X1} + \frac{\hbar^2 k^2}{2m_{x1}} & 0 & 0 & \Omega_{X1} \\ 0 & E_{X1S1} + \frac{\hbar^2 k^2}{2m_{X1S1}} & 0 & \Omega_{X1S1} \\ 0 & 0 & E_{X1S2} + \frac{\hbar^2 k^2}{2m_{X1S2}} & \Omega_{X1S2} \\ \Omega_{X1} & \Omega_{X1S1} & \Omega_{X1S2} & E_C + \frac{\hbar^2 k^2}{2m_C} \end{pmatrix} \quad (1)$$

in which $E_{X1} = 1.554\ eV$, $E_{X1S1} = 1.636\ eV$ and $E_{X1S2} = 1.648\ eV$ are the experimental energy resonances of the exciton and two excited states, respectively, while $E_C = 1.617\ eV$ is the photon energy.

Fig. S7 a-c show the fitting of different reflection spectra as function of the in-plane momentum, decreasing the ReS$_2$ crystal thickness. We extract the Rabi splitting value of 84 meV, 74 meV and 44 meV for 310 nm, 140 nm and 60 nm flakes thickness, respectively.

In order to calculate the exciton and photon fraction of the polariton branches for different thickness, we evaluate the Hopfield coefficients (Fig. S7d-f) from the eigenvectors of the matrix (1). In this way, we consider not only the contribution of the exciton $E_{X1}$ and the photonic mode to the exciton fraction, but also the effect of the two excited states $E_{X1S1}$ and $E_{X1S2}$. Note that in Fig. S7a, the exciton energy (dashed yellow line around 798 nm) is between the longitudinal (L) and transversal (T) modes generated in the ReS$_2$ crystals which are clearly evident in the experimental dispersion spectra for the thickness of d=310nm in Fig. 2b.

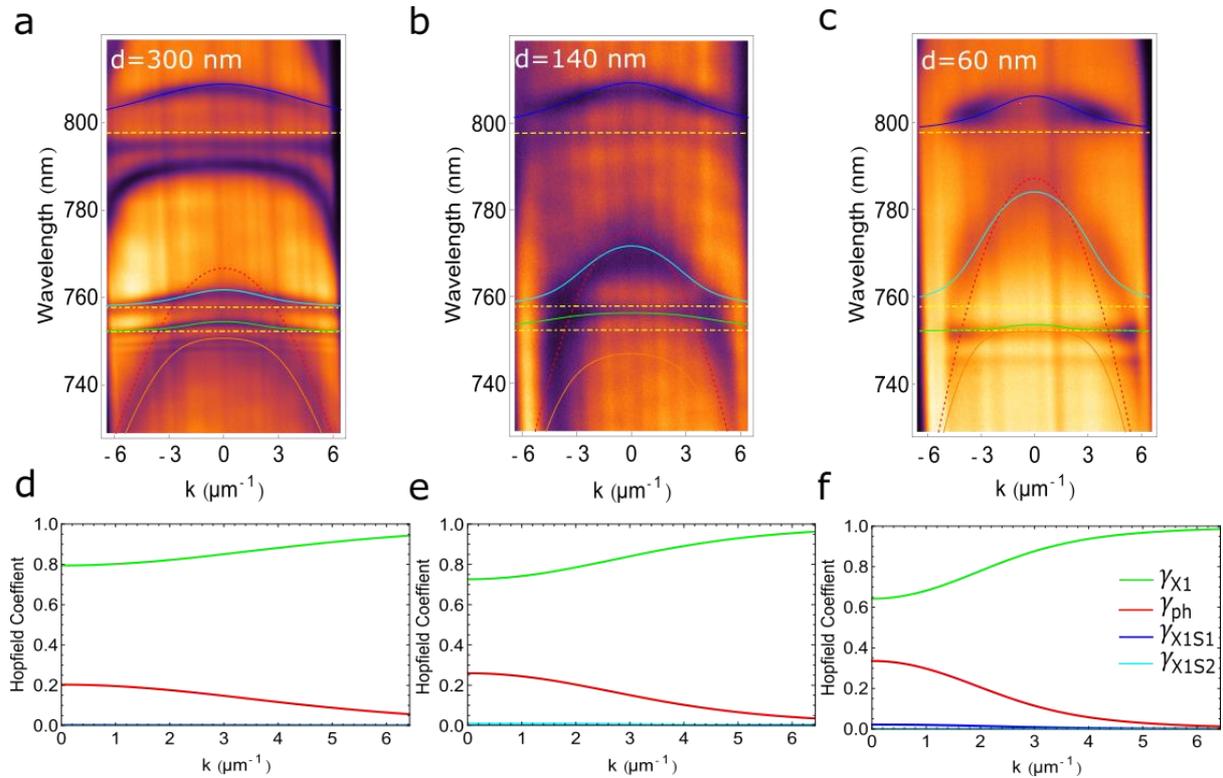

*Figure S7. (a-c) Different polariton branches for ReS$_2$ crystals with thickness of d = 310 nm (a), d = 140 nm (b), and d = 60 nm (c), obtained by assuming the coupling between three exciton states (shown with the yellow dashed lines, $E_{X1}$, $E_{X1S1}$ and $E_{X1S2}$) with only one photonic mode (red dotted lines), which results in a lower polariton branch (blue lines), two middle polariton branches (cyan and green lines) and an upper polariton branch (orange line). (d-f) The Hopfield coefficient for the lower polariton branch for the considered ReS$_2$ thickness.*

**Rydberg States**

The reflectance spectra of thicker ReS$_2$ crystals show peaks at higher energy, in addition to the transitions of the fundamental excitons. These states are excited states of both the excitons and their energies follow the usual hydrogenic Rydberg series of energy levels of 3D excitonic states (En = Ry*/n$^2$).

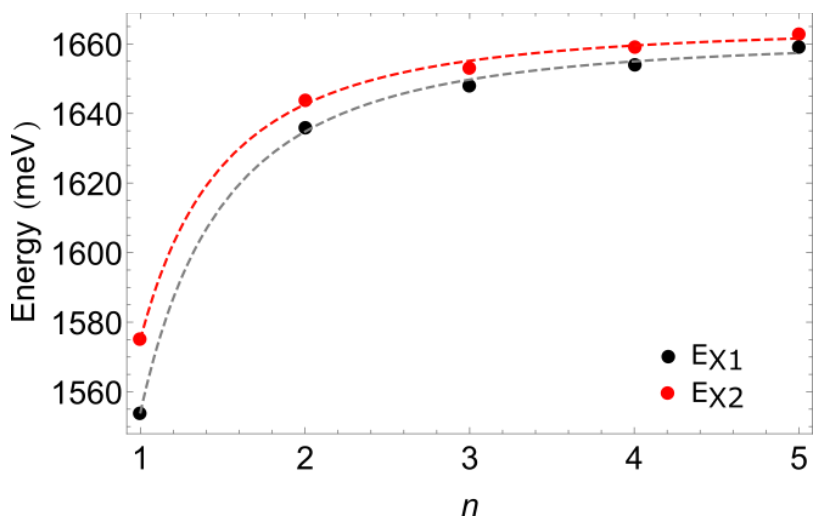

*Figure S8. Experimental data (dots) and theoretical prediction of hydrogenic Rydberg series (dashed lines) of energy exciton states as a function of the quantum number n, for the $E_{X1}$ (black) and $E_{X2}$ (red) excitons.*

The formula used to extract the binding energy of the excited states describing the three dimensional Wannier excitons in inorganic semiconductor[3] is:

$$E_b^{(n)} = E_g - Ry^*/n^2$$

where $E_b^{(n)}$ is the binding energy at the *n*th excitonic state, $E_g$ is the energy gap of the ReS$_2$, $Ry^*$ is the effective Rydberg constant and *n* is the number of exciton state. The fitting parameters extracted for the two Rydberg series associated with the two excitons (for the two polarizations) are reported in Tab. S2:

|  | $E_g$ *(meV)* | *Ry (meV)* |
|---|---|---|
| *Polarization // to b-axis (V)* | 1661 | 107 |
| *Polarization ⊥ to b-axis (H)* | 1665 | 90 |

*Table S2. Fitting parameters extracted for the excited state of the two excitons $E_{x1}$ and $E_{x2}$*